\documentclass[fleqn,12pt]{article}
\usepackage{amsmath}
\usepackage{amsfonts}
\usepackage{latexsym}
\usepackage{amsmath}
\usepackage{rotate,epsfig}
\usepackage{lscape}
\usepackage{color}
\usepackage{graphicx}
\newcommand{\re}[1]{\mbox{${\rm (\ref{#1})}$}}

\headheight\baselineskip
\advance\textwidth -2 in
\advance\textheight-2in \footskip2\baselineskip
\label{sec.sub.gdn}

\date{}

\begin{document}

\date{}

\begin{center}

\large{{\bf Estimation of $P(X>Y)$ for Weibull distribution based on
hybrid censored samples}}
\date{}
\vspace*{0.5 cm}

\normalsize{\bf A. Asgharzadeh}$^{{\rm a}}$\footnote{ {\em E-mail addresses:}
{\it a.asgharzadeh@umz.ac.ir} (A. Asgharzadeh), {\it
m.kazemie64@yahoo.com} (M. Kazemi). }, {\bf M. Kazemi}$^{{\rm a}}$, {\bf D. Kundu}$^{{\rm b}}$\\
\vspace*{0.5 cm} $^{{\rm a}}${\it Department of Statistics, Faculty
of Mathematical Sciences, University of Mazandaran, P. O. Box 47416-1407, Babolsar, Iran}\\
$^{{\rm b}}${\it Department of Mathematics, Indian Institute of
Technology, Kanpour, India}\\

\end{center}

\begin{abstract}
A Hybrid censoring scheme is mixture of Type-I and Type-II censoring
schemes. Based on hybrid censored samples, this paper deals with
the inference on $R = P(X>Y)$, when
X and Y are two independent Weibull distributions with different
scale parameters, but having the same shape parameter. The maximum
likelihood estimator (MLE), and the approximate MLE (AMLE) of R are
obtained. The asymptotic distribution of the maximum likelihood
estimator of $R$ is obtained. Based on the asymptotic distribution,
the confidence interval of $R$ can be derived. Two bootstrap
confidence intervals are also proposed. We consider the Bayesian
estimate of $R$, and propose the corresponding credible interval for
$R$. Monte Carlo simulations are performed to compare the different
proposed methods. Analysis of a real data set has also been
presented for illustrative purposes.
\end{abstract}


\section{ Introduction}

In the context of reliability, the stress-strength model describes
the life of a component which has a random strength $  X$ and is
subjected to random stress $ Y $. The component fails at the instant
that the stress applied to it exceeds the strength and the component
will function satisfactorily if the event $\lbrace(x,y)|x>y\rbrace $
occurs. Thus, $R = P (X > Y)$ is a measure of
component reliability. For some application of $R$, see Kotz \textit{et al}. (2003). Many authors have studied the
stress-strength parameter $R$. Among them, Ahmad \textit{et al}. (1997), Awad
\textit{et al}. (1981), Kundu and Gupta (2005, 2006), Adimari and Chiogna
(2006), Baklizi (2008), Raqab \textit{et al}. (2008) and Rezaei (2010).

A mixture of Type-I and Type-II censoring schemes is known as the
hybrid censoring scheme and it can be described as follows. Suppose
$n$ identical units are put on a life test. The lifetimes of the
sample units are independent and identically distributed (i.i.d)
random variables. The test is terminated when a pre-specified number
$r$ out of $n$ units have failed or a pre-determined time $T$, has
been reached. It is also assumed that the failed items are not
replaced. Therefore, in hybrid censoring scheme, the experimental
time and the number of failures will not exceed $T$ and $r$,
respectively. It is clear that Type-I and Type-II censoring schemes
can be
obtained as special cases of hybrid censoring scheme by taking $r=n$ and $%
T\rightarrow \infty $ respectively. Now we describe the data available under the hybrid censoring scheme. Note
that, under the hybrid censoring scheme, it is assumed that $r$ and $T$ are
known in advance. Therefore, under this censoring scheme we have one of the
two following types of observations:%
\begin{equation*}
\begin{array}{l}
\text{Case I}:\{y_{1:n}<y_{2:n}<\cdots <y_{r:n}\}\text{ if }y_{r:n}<T \\
\text{Case II}:\{y_{1:n}<y_{2:n}<\cdots <y_{d:n}\}\text{ if }d<r,\text{ }%
y_{d:n}<T<y_{d+1:n}%
\end{array}%
,
\end{equation*}%
where $y_{1:n}<y_{2:n}<\cdots $ denote the observed ordered failure times of
the experimental units. For further details on hybrid censoring and relevant references, see Epstein (1954), Fairbanks \textit{et al}. (1982), Childs \textit{et al}. (2003), Gupta and Kundu (1998), Kundu (2007), Ebrahimi (1990, 1992).

The two-parameter Weibull distribution denoted by $W(\alpha,\theta)$
has the probability density function (pdf)
\begin{equation}
f(x,\alpha,\theta)=\frac{\alpha}{\theta}x^{\alpha-1}e^{-\frac{x^\alpha}{\theta}},~~~~x>0,~\alpha,~\theta>0,\label{1}
\end{equation}
and the cumulative distribution function (cdf)
\begin{equation}
F(x,\alpha,\theta)=1-e^{-\frac{x^\alpha}{\theta}},~~~~~~x>0,~\alpha,~\theta>0,\label{2}
\end{equation}
Here $\alpha$ is the shape parameter and $\theta$ is the scale
parameter.

Based on complete $X$-sample and $Y$-sample, Kundu and Gupta (2006)
considered the estimation of $R=P(Y<X)$ when $X\sim
W(\alpha,\theta_1)$ and $Y\sim W(\alpha,\theta_2)$ are two
independent Weibull distributions with different scale parameters,
but having the same shape parameter. In this paper, we extend their
results for the case when the samples are hybrid
censored.

The paper is organized as follows: In Section 2, we derive the
maximum likelihood estimator (MLE) of $R$. It is observed that the
MLE can not be obtained in a closed form. We propose an approximate
MLE (AMLE) of $R$ in Section 3, which can be obtained explicitly.
Different confidence intervals are presented in Section 4. Bayesian
solutions are presented in Section 5. Analysis of a real data set as
well a Monte Carlo simulation based comparison of the proposed
methods are performed in Section 6.  Finally in Section 7, we conclude the paper.

\section{Maximum Likelihood Estimator of $R$}

Let $X\sim W(\alpha,\theta_1)$ and $Y\sim W(\alpha,\theta_2)$ be
independent random variables. Then it can be easily seen that
\begin{equation}
R=P(X>Y)=\frac{\theta_1}{\theta_1+\theta_2}.\label{3}
\end{equation}

Our interest is in estimating $R$ based on hybrid censored data on both variables.
To derive the MLE of $R$, first we obtain the MLE's of $\alpha$, $\theta_1$ and $\theta_2$. Suppose $\textbf{X}=(X_{1:n},X_{2:n},\cdots,X_{r_{1}:n})$ is a hybrid censored sample from $W(\alpha,\theta_1)$ with censored scheme $\textbf(r_{1},T_{1})$ and $\textbf{Y}=(Y_{1:m},Y_{2:m},\cdots,Y_{r_{2}:m})$ is a hybrid censored sample from  $W(\alpha,\theta_2)$ with censored scheme $\textbf(r_{2},T_{2})$. For notation simplicity, we will write $(X_{1},X_{2},\cdots,X_{r_{1}})$ for $(X_{1:n},X_{2:n},\cdots,X_{r_{1}:n})$ and $(Y_{1},Y_{2},\cdots,Y_{r_{2}})$ for $(Y_{1:m},Y_{2:m},\cdots,Y_{r_{2}:m})$. Therefore, the likelihood function of $\alpha$, $\theta_1$
and $\theta_2$ is given (see Balakrishnan and Aggarwala (2000)) by
\begin{equation}
L(\alpha,\theta_1,\theta_2)=\left[c_{1}\prod_{i=1}^{r_{1}}f(x_{i})[1-F(u_{1})]^{n-r_{1}}\right]\times
\left[c_{2}\prod_{j=1}^{r_{2}}f(y_{j})[1-F(u_{2})]^{m-r_{2}}\right],\label{4}
\end{equation}
where
\begin{eqnarray*}
c_{1}&=&n(n-1)(n-2)\cdots(n-r_{1}+1),   ~~~~~~~~~~~~~ U_{1}=min(X_{R_{1}},T_{1}),\\
c_{2}&=&m(m-1)(m-2)\cdots(m-r_{2}+1),  ~~~~~~~~~~  U_{2}=min(Y_{R_{2}},T_{2}).
\end{eqnarray*}
and
\begin{eqnarray*}
r_{1}=\sum_{i=1}^{r_{1}}I\left\{{x_{i}\leq u_{1}}\right\},  ~~~~~  r_{2}=\sum_{j=1}^{r_{2}}I\left\{{y_{j}\leq u_{2}}\right\}.
\end{eqnarray*}
\noindent Upon using \re{1} and \re{2}, we immediately have the likelihood function of the observed data as follows:
\begin{eqnarray}
L(data|\alpha,\theta_1,\theta_2)&=&c_1c_2\alpha^{r_{1}+r_{2}}\theta_1^{-r_{1}}\theta_2^{-r_{2}}
\prod_{i=1}^{r_{1}}x_i^{\alpha-1}\prod_{j=1}^{r_{2}}y_j^{\alpha-1}\nonumber\\
& \times &\exp\left\{-\frac{1}{\theta_1}\left[\sum_{i=1}^{r_{1}}x_i^\alpha +(n-r_{1})u_{1}^\alpha\right]-\frac{1}{\theta_2}\left[\sum_{j=1}^{r_{2}}y_j^\alpha +(m-r_{2})u_{2}^\alpha\right]\right\rbrace. ~~~~~\label{5}
\end{eqnarray}
From (5), the log-likelihood function for the hybrid censored data without the multiplicative constant is
\begin{eqnarray}
l(\alpha,\theta_1, \theta_2)& = & \hspace{-.25cm} (r_{1}+r_{2})\ln (\alpha)-r_{1}\ln(\theta_1)-r_{2}\ln(\theta_2)+(\alpha-1)\times \left[\sum_{i=1}^{r_{1}}\ln(x_i)+\sum_{j=1}^{r_{2}}\ln(y_j)\right]\nonumber\\
&-&\frac{1}{\theta_1}\left[\sum_{i=1}^{r_{1}}x_i^\alpha +(n-r_{1})u_{1}^\alpha\right]-\frac{1}{\theta_2}\left[\sum_{j=1}^{r_{2}}y_j^\alpha +(m-r_{2})u_{2}^\alpha\right].~~~~\label{6}
\end{eqnarray}

The MLEs of $\alpha$, $\theta_1$ and $\theta_2$, say $\widehat\alpha$, $\widehat\theta_1$ and $\widehat\theta_2$
respectively, can be obtained as the solution of
\begin{eqnarray}
\frac{\partial l}{\partial\alpha}&=&\frac{r_{1}+r_{2}}{\alpha}+\left[\sum_{i=1}^{r_{1}}\ln(x_i)+\sum_{j=1}^{r_{2}}\ln(y_j)\right]
-\frac{1}{\theta_1}\left[\sum_{i=1}^{r_{1}}x_i^\alpha \ln(x_i)+(n-r_{1})u_{1}^\alpha\ln(u_{1})\right]\nonumber\\
&-&\frac{1}{\theta_2}\left[\sum_{j=1}^{r_{2}}y_j^\alpha\ln(y_j)+(m-r_{2})u_{2}^\alpha\ln(u_{2})\right]=0,
\label{5}
\end{eqnarray}
\begin{eqnarray}
\frac{\partial l}{\partial\theta_1}=-\frac{r_1}{\theta_1}+\frac{1}{\theta_1^2}\left[\sum_{i=1}^{r_{1}}x_i^\alpha +(n-r_{1})u_{1}^\alpha\right]=0, \label{6}
\end{eqnarray}
\begin{eqnarray}
\frac{\partial l}{\partial\theta_2}=-\frac{r_2}{\theta_2}+\frac{1}{\theta_2^2}\left[\sum_{j=1}^{r_{2}}y_j^\alpha +(m-r_{2})u_{2}^\alpha\right]=0.\label{7}
\end{eqnarray}

\noindent From (7) and (8), we obtain
\begin{equation}
\widehat{\theta}_{1}(\alpha)=\frac{1}{r_1}\left[\sum_{i=1}^{r_1}x_i^\alpha
+(n-r_{1})u_{1}^\alpha\right],\;\;\mbox{and}\;\;
\widehat{\theta}_{2}(\alpha)=\frac{1}{r_2}\left[\sum_{j=1}^{r_2}y_j^\alpha
+(m-r_{2})u_{2}^\alpha\right].\label{8}
\end{equation}

Substituting the expressions of $\widehat{\theta}_1(\alpha)$ and
$\widehat{\theta}_2(\alpha)$ into \re{5}, $\widehat{\alpha}$ can be obtained as a fixed point solution of the
following equation:
\begin{equation}
\hspace{5cm}k(\alpha)=\alpha.\label{11}
\end{equation}
Here
$$
k(\alpha)=\frac{r_1+r_2}{U(\alpha) + V(\alpha) - W},
$$
where
$$
U(\alpha) = \frac{r_1 \left [\sum_{i=1}^{r_{1}}x_i^\alpha \ln(x_i)+(n-r_{1})u_{1}^\alpha\ln(u_{1}) \right ]}{\sum_{i=1}^{r_{1}}x_i^\alpha +(n-r_{1})u_{1}^\alpha},
$$
$$
V(\alpha) =  \frac{r_2\left  [\sum_{j=1}^{r_{2}}y_j^\alpha \ln(y_j)+(m-r_{2})u_{2}^\alpha\ln(u_{2}) \right ]}{\sum_{j=1}^{r_{2}}y_j^\alpha +(m-r_{2})u_{2}^\alpha}
$$
and
$$
W = \sum_{i=1}^{r_1}\ln(x_i)+\sum_{j=1}^{r_2}\ln(y_j).
$$

A simple iterative procedure $k(\alpha^{(j)})=\alpha^{(j+1)}$, where
$\alpha^{(j)}$ is the j-th iterate, can be used to find the solution
of (11). The iterative procedure should be stopped when the absolute difference between
 $ \alpha^{(j)}$ and $\alpha^{(j+1)}$  is sufficiently small. Once we obtain $\widehat{\alpha}_{ML}$, the MLE of
$\theta_1$ and $\theta_2$, can be deduced from \re{8} as
$\widehat{\theta}_{1ML}=\widehat{\theta}_1(\widehat{\alpha}_{ML})$
and
$\widehat{\theta}_{2ML}=\widehat{\theta}_2(\widehat{\alpha}_{ML})$.
Therefore, we compute the MLE of $R$ as
\begin{eqnarray}
\widehat{R}_{ML}=\frac{\frac{1}{r_1}\sum_{i=1}^{r_1}x_i^{\widehat\alpha_{ML}}+(n-r_1)u_1^{\widehat\alpha_{ML}}}
{{\frac{1}{r_1}\sum_{i=1}^{r_1}x_i^{\widehat\alpha_{ML}}+(n-r_1)u_1^{\widehat\alpha_{ML}}}+\frac{1}{r_2}\sum_{j=1}^{r_2}y_j^{\widehat\alpha_{ML}}+(m-r_2)u_2^{\widehat\alpha_{ML}}}.
\label{12}
\end{eqnarray}

Here the maximum likelihood approach does not give an explicit
estimator for $\alpha$ and hence for $R$, based on a hybrid censored
sample. In the next section, we propose the approximate maximum
likelihood estimates which have explicit forms.

\section{Approximate Maximum Likelihood Estimator of $R$}

In this section, the approximate maximum likelihood  method is used
to estimate the stress-strength parameter $R$. It is based on the
fact that if the random variable $X$ has $W(\alpha, \theta)$, then
 $V=\ln(X)$, has the extreme value distribution with pdf as
\begin{equation}
f(v;\mu,\sigma)=\frac{1}{\sigma}e^{\frac{v-\mu}{\sigma}-e^{\frac{v-\mu}{\sigma}}},\;\;-\infty<v<\infty,\label{11}
\end{equation}
\noindent where $\mu=\dfrac{1}{\alpha}\ln \theta$ and
$\sigma=\dfrac{1}{\alpha}$. The density function \re{11} is
known as the density function of an extreme value distribution, with
location, and scale parameters as $\mu$ and $\sigma$ respectively.
The standard extreme value distribution has the pdf and cdf as
$$
g(v)=e^{v-e^{v}},\hspace{3cm} G(v)=1-e^{-e^v}.
$$

Suppose $X_{1}<X_{2}<\cdots<X_{r_1}$ is ahybrid censored sample from $W(\alpha,\theta_1)$ with censored scheme $(r_{1},T_{1})$ and $Y_{1}<Y_{2}<\cdots<Y_{r_2}$ is a hybrid censored sample from $W(\alpha,\theta_2)$ with censored scheme $(r_{2},T_{2})$. Let us use the following notations: $T_i=\ln(X_i),\ Z_i=\dfrac{T_i-\mu_1}{\sigma},\ i=1,\cdots,r_1$ and $S_j=\ln(Y_j),\ W_j=\dfrac{S_j-\mu_2}{\sigma}, \ j=1,\cdots,r_2$, where $\mu_1=\dfrac{1}{\alpha}\ln \theta_1$ , $\mu_2=\dfrac{1}{\alpha}\ln
\theta_2$ and $\sigma=\dfrac{1}{\alpha}$.

\noindent The log-likelihood function of the observed data $T_1,\cdots,T_{r_1}$ and $S_1,\cdots,S_{r_2}$ is
\begin{eqnarray}
l^*(\mu_1,\mu_2,\sigma)&\propto& -(r_1+r_2)\ln \sigma+\sum_{i=1}^{r_1}\ln(g(z_i))+(n-r_1)\ln(1-G(u_1^*))\nonumber\\
&+&\sum_{j=1}^{r_2}\ln(g(w_j))+(m-r_2)\ln(1-G(u_2^*)),\label{12}
\end{eqnarray}

\noindent where $ u_1^*=\dfrac{\ln u_1-\mu_1}{\sigma},\ u_2^*=\dfrac{\ln u_2-\mu_2}{\sigma}.$ Differentiating \re{12} with respect to $\mu_1$, $\mu_2$ and $\sigma$, we obtain the likelihood equations as
\begin{eqnarray}
\frac{\partial l^*}{\partial\mu_1}&=&-\frac{1}{\sigma}\sum_{i=1}^{r_1}\frac{g'(z_i)}{g(z_i)}
+\frac{1}{\sigma}(n-r_1)\frac{g(u_1^*)}{1-G(u_1*)}= 0,\label{13}\\
\frac{\partial l^*}{\partial\mu_2}&=&-\frac{1}{\sigma}\sum_{j=1}^{r_2}\frac{g'(w_j)}{g(w_j)}+\frac{1}{\sigma}(m-r_2)
\frac{g(u_2^*)}{1-G(u_2^*)}= 0,\label{14}\\
\frac{\partial l^*}{\partial\sigma}&=&-\frac{r_1+r_2}{\sigma}-\frac{1}{\sigma}\sum_{i=1}^{r_1}z_i\frac{g'(z_i)}{g(z_i)}+\frac{1}{\sigma}(n-r_1)u_1^*\frac{g(u_1^*)}{1-G(u_1^*)}\nonumber\\
&-&\frac{1}{\sigma}\sum_{j=1}^{r_2}w_j\frac{g'(w_j)}{g(w_j)}
+\frac{1}{\sigma}(m-r_2)u_2^*\frac{g(u_2^*)}{1-G(u_2^*)}=0.\label{15}
\end{eqnarray}

It is observed that the likelihood equations are not linear and do
not admit explicit solutions. We approximate the terms
$p(z_i)=\dfrac{g'(z_i)}{g(z_i)}$ and
$q(u_1^*)=\dfrac{g(u_1^*)}{1-G(u_1^*)}$ by expanding  in Taylor
series as follows. Suppose $p_i=\frac{i}{n+1}$, $q_i=1-p_i$  and
$p_{r_{i}}^*=\dfrac{p_{r_i}+p_{r_{i}+1}}{2},q_{r_i}^*=1-p_{r_i}^* $.
We expand the function $p(z_i)$ around $
G^{-1}(p_i)=\ln(-\ln(q_{i}))=\mu_i$. Further, we also expand the
term $q(u_1^*)$ around $G^{-1}(p_{R_{1}})=\mu_{R_{1}}$ if
$u_1=x_{R_1}$. If  $u_1=T_1$,  expand $q(u_1^*)$ around the point
$G^{-1}(p_{r_1}^*)=\mu_{r_1}^*$.

Similarly, We approximate the function $\bar{p}{(w_j)}=\dfrac{g'(w_j)}{g(w_j)}$ around $ G^{-1}(p_j)=\mu_j$. we also expand the term $\bar{q}(u_2^*)=\dfrac{g(u_2^*)}{1-G(u_2^*)}$ around $G^{-1}(p_{R_{2}})=\mu_{R_2}$ if $u_2=x_{R_2}$ and If  $u_2=T_2$,  we expand $\bar{q}(u_2^*)$ around the point $G^{-1}(p_{r_2}^*)=\mu_{r_2}^*$.
Now, let $u_1=x_{R_1}$ and $u_2=x_{R_2}$. Then $r_1=R_1,r_2=R_2$ and $u_1^*=z_{R_1}, u_2^*=w_{R_2}$.\

Considering only the first order derivatives and neglecting the higher order derivatives we get
\begin{eqnarray*}
 p(z_{i}) &\approx & \alpha_{i}+\beta_{i}z_{i},\hspace{4cm}\bar{p}(w_{j}) \approx  {\alpha}_{j}+{\beta}_{j}w_{j},\\
q(z_{R_1}) &\approx & \ 1- \alpha_{R_1}-\beta_{R_1}z_{R1},\hspace{2.1cm}\bar{q}(w_{R_2})\approx \ 1- \alpha_{R_2}-\beta_{R_2}z_{R_2},
\end{eqnarray*}
where
\begin{eqnarray*}
\alpha_i &=& 1+\ln q_i(1-\ln(-\ln q_i)), \hspace{1cm}\beta_i=\ln (q_i).
\end{eqnarray*}
\noindent Therefore, \re{13}, \re{14}, and \re{15} can be approximated respectively
as
\begin{eqnarray}
\frac{\partial l^*}{\partial\mu_1}&=&-\frac{1}{\sigma}\left[\sum_{i=1}^{R_1}(\alpha_i+\beta_i\ z_i)
-(n-R_1)(1-\alpha_{R_1}-\beta_ {R_1}\ z_{R_1})\right]= 0,\label{17}\\
\frac{\partial l^*}{\partial\mu_2}&=&-\frac{1}{\sigma}\left[\sum_{j=1}^{R_2}(\alpha_j+\beta_j\ w_j)
-(m-R_2)(1-\alpha_{R_2}-\beta _{R_2}\ w_{R_2})\right]= 0,\label{18}\\
\frac{\partial l^*}{\partial\sigma}&=&-\frac{2(R_1+R_2)}{\sigma}-\frac{1}{\sigma}\left[\sum_{i=1}^{R_1}
z_i(\alpha_i+\beta_i\ z_i)-(n-R_1)(1-\alpha_{R_1}-\beta _{R_1}\ z_{R_1})\right]\nonumber\\
&-&\frac{1}{\sigma}\left[\sum_{j=1}^{R_2}w_j(\alpha_j+\beta_j\ w_j)-(m-R_2)(1-\alpha_{R_2}-\beta_{R_2}\ w_{R_2})\right]=0.\label{19}
\end{eqnarray}

\noindent If we denote $\tilde{\mu}_1$, $\tilde{\mu}_2$ and $\tilde{\sigma}_1$ as the solutions of  (18), (19) and  (20) respectively, then observe that
\[
\tilde{\mu}_1=A_1+B_1\tilde{\sigma},\;\;\tilde{\mu}_2=A_2+B_2\tilde{\sigma},\;\;\mbox{and}\;\;
\tilde{\sigma}=\frac{-D+\sqrt{D^2-4(R_1+R_2)E}}{2(R_1+R_2)},
\]
\noindent where

\begin{eqnarray*}
A_1=\frac{\sum_{i=1}^{R_1}\beta_i t_i+(n-R_1)\beta_{R_1}t_{R_1}}{\sum_{i=1}^{R_1}\beta_i+(n-R_1)\beta_{R_1}},&& B_1=\frac{\sum_{i=1}^{R_1}\alpha_i -(n-R_1)(1-\alpha_{R_1})}{\sum_{i=1}^{R_1}\beta_i+(n-R_1)\beta_{R_1}},\
\end{eqnarray*}

\begin{eqnarray*}
A_2=\frac{\sum_{j=1}^{R_2}\beta_j s_j+(m-R_2)\beta_{R_2}s_{R_2}}{\sum_{j=1}^{R_2}\beta_j+(m-R_2 )\beta_{R_2}},
&&B_2=\frac{\sum_{j=1}^{R_2}\alpha_j -(m-R_2)(1-\alpha_{R_2})}{\sum_{j=1}^{R_2}\beta_j+(m-R_2)\beta_{R_2}},
\end{eqnarray*}

\begin{eqnarray*}
D&=&\sum_{i=1}^{R_1}\alpha_i(\ t_i -3A_1)-(n-R_1)(1-\alpha_{R_1}) (t_{R_1}-3A_1)+2A_1B_1
\sum_{i=1}^{R_1}\beta_i +2A_1 B_1(n-R_1)\beta_{R_1}\\
 &+&\sum_{j=1}^{R_2}\alpha_j(\ s_j -3A_2)-(m-R_2)(1-\alpha_{R_2})(s_{R_2}-3A_2)+2A_2B_2\sum_{j=1}^{R_2}\beta_i+2A_2 B_2(m-R_2)\beta_{R_2},\\
\\
E&=&\sum_{i=1}^{R_1}\beta_i t_i(t_i-A_1)+(n-R_1)\beta_{R_1}t_{R_1}^2-(n-R_1)A_1\beta_{R_1}t_{R_1}
+\sum_{j=1}^{R_2}\beta_j s_j(s_j-A_2) \hspace{1cm}\\
&+&(m-R_2)\beta_{R_2}s_{R_2}^2 -(m-R_2)A_2\beta_{R_2}s_{R_2}.
\end{eqnarray*}

Once $\tilde{\sigma}$ is obtained, $\tilde{\mu}_1$  and $\tilde{\mu}_2$ can be obtained immediately. Therefore, the approximate MLE of $R$ is given by
\[
\tilde{R}=\frac{\tilde{\theta}_1}{\tilde{\theta}_1+\tilde{\theta}_2},
\]
\noindent where
\[
\tilde{\theta}_1=\exp{\left(\frac{1}{\tilde{\sigma}}(A_1+B_1\tilde{\sigma})\right)},\;\;
\mbox{and}\;\;\tilde{\theta}_2=\exp{\left(\frac{1}{\tilde{\sigma}}(A_2+B_2\tilde{\sigma})\right)}.
\]

\noindent Similar procedure as above can be used to obtain AMLE of $R$ for other cases (See the Appendix).

\section{Different Confidence Intervals of $R$}

Based on asymptotic behavior of $R$, we present an asymptotic confidence interval of $R$. We further, propose two confidence intervals based on the non-parametric bootstrap method.

\subsection{Asymptotic Confidence Intervals of $R$}

In this subsection, we obtain the asymptotic variances and covariances
of the MLEs, $\widehat{\theta_1}, \widehat{\theta_2}$ and
$\widehat{\alpha}$ by entries of the inverse of the observed Fisher
information matrix ${\bf I(\theta})$ where
$\theta=(\alpha,\theta_1,\theta_2)$.
 From the log-likelihood function in (6), we obtain the observed Fisher
information matrix of $\theta $ as

{\large
$$
I(\theta)=-\left(\begin{array}{ccc}
\frac{\partial^2 l}{\partial\alpha^2}& \frac{\partial^2 l}{\partial\alpha\partial\theta_1}
&\frac{\partial^2 l}{\partial\alpha\partial\theta_2}\\
\frac{\partial^2 l}{\partial\theta_1\partial\alpha}&\frac{\partial^2 l}{\partial\theta_1^2}&
\frac{\partial^2 l}{\partial\theta_1\partial\theta_2}\\
\frac{\partial^2 l}{\partial\theta_2\partial\alpha}&\frac{\partial^2 l}{\partial\theta_2\partial\theta_1}&
\frac{\partial^2 l}{\partial\theta_2^2}
\end{array}\right)=
\left(\begin{array}{ccc}
I_{11}& I_{12} & I_{13} \\
I_{21} & I_{22} & I_{23} \\
I_{31} & I_{32} & I_{33} \\
\end{array}  \right),
$$
}

where
\begin{eqnarray*}
-I_{11}&=&-\frac{r_{1}+r_{2}}{\alpha^2}-\frac{1}{\theta_1}\bigg [\sum_{i=1}^{r_{1}}x_i^\alpha[\ln(x_i)]^2  +
(n-r_1)u_{1}^{\alpha}[\ln(u_1)]^2\bigg] \\
&-&\frac{1}{\theta_2}\bigg[\sum_{j=1}^{r_{2}}y_j^\alpha[\ln(y_j)]^2+(m-r_2)u_{2}^{\alpha}[\ln(u_{2})]^2\bigg],\\
-I_{22}&=&\frac{r_{1}}{\theta_1^2}-\frac{2}{\theta_1^3}\Bigg[\sum_{i=1}^{r_{1}}x_i^\alpha
+(n-r_{1})u_{1}^{\alpha}\Bigg],\\
-I_{33}&=&\frac{r_{2}}{\theta_2^2}-\frac{2}{\theta_2^3}\Bigg[\sum_{j=1}^{r_{2}}y_j^\alpha
+(m-r_{2})u_{2}^{\alpha} \Bigg],\\
-I_{12}&=&\frac{1}{\theta_1^2}\bigg[ \sum_{i=1}^{r_{1}}x_i^\alpha\ln(x_i)+(n-r_{1})u_{1}^{\alpha}\ln (u_{1})\bigg]=-I_{21},\\
-I_{13}&=&\frac{1}{\theta_2^2}\bigg[ \sum_{j=1}^{r_{2}}y_j^\alpha\ln(y_j)+(m-r_{2})u_{2}^{\alpha}\ln (u_{2})\bigg]=-I_{31},\\
I_{23}&=&I_{32}=0,
\end{eqnarray*}

Now, the asymptotic variance-covariance matrix, $A=[a_{ij}]$, for
the MLE's is obtained by inverting the Fisher information matrix,
i.e. $A=I(\theta)^{-1} =[I_{ij} ]^{-1}$. Therefore

$$
{\bf A}=\frac{1}{u}
\left(\begin{array}{ccc}I_{22}I_{33} & -I_{12}I_{33} & -I_{22}I_{13} \\
 -I_{21}I_{33} & I_{11}I_{33}-I_{13}I_{31} & I_{21}I_{13} \\
 -I_{22}I_{31} & I_{12}I_{31} & I_{11}I_{22}-I_{12}I_{21} \\
\end{array}\right),
$$

\noindent where
\begin{center}
$
u=I_{11}I_{22}I_{33}-I_{12}I_{21}I_{33}-I_{13}I_{31}I_{22}.
$
\end{center}
Now, the variance of $\widehat{R}$ (say $B$) can obtained also using
delta method as follows. We have
$\widehat{R}=g(\widehat{\alpha},\widehat{\theta_1},\widehat{\theta_2}), $
where
$$
g(\alpha,\theta_1,\theta_2)=\dfrac{\theta_1 }{\theta_1+\theta_2}.
$$
Therefore, $B=\textbf{b}^t{\bf A}\textbf{b}$, where
$$
\textbf{b}=\left(%
\begin{array}{c}
 \frac{\partial g}{\mathbf{\partial \alpha}} \\
  \frac{\partial
g}{\mathbf{\partial \theta_1}} \\
  \frac{\partial g}{\mathbf{\partial \theta_2 }} \\
\end{array}
\right)=\frac{1}{(\theta_1+\theta_2)^2}\left(\begin{array}{c} 0 \\ \theta_2 \\ -\theta_1 \\
\end{array}\right),
$$
It is easy to show that
$$
B=\textbf{b}'{\bf A}\textbf{b}=
\dfrac{1}{u(\theta_1+\theta_2)^4}\left[(I_{11}I_{22}-I_{12}^2)
\theta_1^2-2 I_{12}I_{13}\theta_1\theta_2+(I_{11}I_{33}-I_{13}^2)\theta_2^2\right].
$$

 To compute the confidence interval of R, the variance B needs to be estimated.
We recommend using the emprical Fisher information matrix, and the MLE estimates of $\alpha$, $\theta_1$ and $\theta_2$, to estimate $ B $, which is very convenient. As a consequence of that, a $100(1-\gamma)\%$ asymptotic
confidence intervals for $R$ can be given by

$$
(\widehat{R}-z_{1-\frac{\gamma}{2}} \sqrt{\widehat{B}},
\widehat{R}+z_{1-\frac{\gamma}{2}} \sqrt{\widehat{B}} ),
$$
\noindent where $z_{\gamma}$ is $100\gamma$-th percentile of $N(0, 1)$. The confidence interval of $R$ by using the asymptotic distribution of the AMLE of $R$ can be obtained similarly by replacing $\alpha$, $\theta_1$ and $\theta_2$ in $B$ by their respective AMLE's.

\vspace{1ex}
 It is of interest to observe that when the shape parameter $\alpha$ is known, the MLE of $R$ can be obtained explicitly as
\begin{equation}
\hat{R}_{ML}= \frac{S_1({\bf x})}{S_1({\bf x})+\frac{r_1}{r_2} S_2({\bf y})},\label{19}
\end{equation}
\noindent where $S_1({\bf x})=\sum_{i=1}^{r_1}x_i^{\alpha}+(n-r_1) u_{1}^{\alpha}$ and $S_2({\bf y})=\sum_{j=1}^{r_2}y_j^{\alpha}+(m-r_2) u_{2}^{\alpha}$.

\vspace{1ex}

\subsection{Bootstrap Confidence Intervals}

In this section,  we propose two confidence intervals based on the
non-parametric bootstrap methods: (i) percentile bootstrap method
(we call it Boot-p) based on the idea of Efron (1982), and (ii)
bootstrap-t method (we refer to it as Boot-t) based on the idea of
Hall (1988). We illustrate briefly how to estimate confidence
intervalss of $R$ using both methods.

\noindent \textbf{(i) Boot-p method}
\begin{enumerate}
\item[1.] Generate a bootstrap sample of size $ r_{1}$, $\{x_1^*,\cdots,x_{r_{1}}^*\}$ from $\{x_1,\cdots,x_{r_{1}}\}$, and generate a bootstrap sample
of size $r_{2}$, $\{y_1^*,\cdots,y_{r_{2}}^*\}$ from $\{y_1,\cdots,y_{r_{2}}\}$.
Based on $\{x_1^*,\cdots,x_{r_{1}}^*\}$ and $\{y_1^*,\cdots,y_{r_{2}}^*\}$,
compute the bootstrap estimate of $R$ say $\widehat{R}^*$ using
(12).
\item[2.] Repeat 1 NBOOT times.
\item[3.] Let $H_1(x)=P(\widehat{R}^*\leq x)$ be the cumulative distribution
function of $\widehat{R}^*$. Define $\widehat{R}_{Bp}(x)=H_1^{-1}(x)$ for a given $x$. The approximate $100(1-\gamma)\%$ confidence
interval of $R$ is given by
$$
\left(\widehat{R}_{Bp}(\frac{\gamma}{2})\ ,\ \widehat{R}_{Bp}(1-\frac{\gamma}{2})\right)
$$
\end{enumerate}
\textbf{(ii) Boot-t method}
\begin{enumerate}
\item[1.] From the sample $\{x_1,\cdots,x_{r_{1}}\}$ and $\{y_1,\cdots,y_{r_{2}}\}$, compute $\widehat{R}$.
\item[2.] Same as in Boot-p method, first generate bootstrap sample $\{{x_1}^*,\cdots,{x_{r_{1}}}^*\},$
$ \{{y_1}^*,\cdots,{y_{r_{2}}}^*\}$ and then compute
$\widehat{R}^*$, the bootstrap estimate of $R$. Also, compute
the statistic
$$
T^*=\frac{\widehat{R}^*-\widehat{R}}{\sqrt{Var(\widehat{R}^*)}}.
$$
\item[3.] Repeat 1 and 2 NBOOT times.
\item[4.] Let $H_2(x)=P(T^*\leq x)$ be the cumulative distribution
function of $T^*$. For a given $x$ define $\widehat{R}_{Bt}(x)=\widehat{R}+H_2^{-1}(x)\sqrt{Var(\widehat{R})}$. The approximate $100(1-\gamma)\%$ confidence intervals of $R$ is given by
$$
\left(\widehat{R}_{Bt}(\frac{\gamma}{2})\ ,\
\widehat{R}_{Bt}(1-\frac{\gamma}{2})\right).
$$
\end{enumerate}

\section{Bayes Estimation of $R$}

In this section, we obtain the Bayes estimation of $R$ under
assumption that the shape parameter $\alpha$ and scale parameters
$\theta_1$ and $\theta_2$ are independent random variables. it is
also assumed that the prior density of $\theta_j$ is the inverse
gamma $ IG(a_j,b_j)$, $j=1,2$ with density function
$$
\pi_{j}(\theta_j)=\pi(\theta_j|a_j,b_j)=e^{-\frac{b_j}{\theta_j}}\frac{\theta_j^{-(a_j+1)}b_j^{a_j}}{\Gamma(a_j)},
$$
and $\alpha$ has the gamma $ G(a_3,b_3)$ with
density function
$$
\pi_{3}(\alpha)=\pi(\alpha|a_3,b_3)=e^{-b_3\alpha}\frac{\alpha^{a_3-1}b_3^{a_3}}{\Gamma(a_3)}.
$$
The likelihood function of the observed data is
\begin{eqnarray}
L(data|\alpha,\theta_1,\theta_2)& \propto &\alpha^{r_{1}+r_{2}}\theta_1^{-r_{1}}\theta_2^{-r_{2}}
\prod_{i=1}^{r_{1}}x_i^{\alpha-1}\prod_{j=1}^{r_{2}}y_j^{\alpha-1}\nonumber\\
& \times &\exp\left\{-\frac{1}{\theta_1}\left[\sum_{i=1}^{r_{1}}x_i^\alpha +(n-r_{1})u_{1}^\alpha\right]-\frac{1}{\theta_2}\left[\sum_{j=1}^{r_{2}}y_j^\alpha +(m-r_{2})u_{2}^\alpha\right]\right\rbrace. ~~~~~
\end{eqnarray}
The joint density of the $ \alpha,\theta_1,\theta_2 $ and data can be
obtained as
$$
L(\alpha,\theta_1,\theta_2, data)=L(data|\alpha,\theta_1,\theta_2)\times \pi_1(\theta_1)\times \pi_2(\theta_2)
\times \pi_3(\alpha).$$
Therefore the joint posterior density of $\alpha$, $\theta_1$ and $\theta_2$ given the data is
\begin{equation}
L(\alpha,\theta_1,\theta_2|data)=\frac{L(data|\alpha,\theta_1,\theta_2)\; \pi_1(\theta_1)\;\pi_2(\theta_2)\;\pi_3(\alpha)}
{\int_0^\infty\int_0^\infty\int_0^\infty L(data|\alpha,\theta_1,\theta_2)\pi_1(\theta_1)\;\pi_2(\theta_2)\;\pi_3(\alpha) \;d \theta_1 \;d\theta_2 \;d\alpha}. \label{20}
\end{equation}
\noindent From \re{20}, it is observed that the Bayes estimate can
not be obtained in a closed form. Therefore,  we opt for stochastic
simulation procedures, namely, the Gibbs and Metropolis samplers
(Gilks et al., 1995) to generate samples from the posterior
distributions and then compute the Bayes estimate of $R$ and the
corresponding credible interval of $R$. The posterior pdfs of
$\theta_1$ and $\theta_2$ are as follows:
\[
\theta_1|\alpha,\theta_2,data\sim
IG\left(r_{1}+a_1,b_1+\sum_{i=1}^{r_1}x_i^\alpha + (n-r_{1})u_{1}^{\alpha}\right),
\]
\[
\theta_2|\alpha,\theta_1,data\sim
IG\left(r_{2}+a_2,b_2+\sum_{j=1}^{r_2}y_j^\alpha + (m-r_{2})u_{2}^{\alpha}\right),
\]
and
\begin{eqnarray*}
f(\alpha|\theta_1,\theta_2,data)&\propto&\alpha^{r_{1}+r_{2}+a_3-1}\prod_{i=1}^{r_1}x_i^{\alpha-1}\prod_{j=1}^{r_2}y_j^{\alpha-1}\\
&\times&\exp\left\{-b_3\alpha-\frac{1}{\theta_1}\Bigg(\sum_{i=1}^{r_1}x_i^\alpha + (n-r_{1})u_{1}^{\alpha}\Bigg)-\frac{1}{\theta_2}\Bigg(\sum_{j=1}^{r_2}y_j^\alpha + (m-r_{2})u_{2}^{\alpha}\Bigg)\right\}.
\end{eqnarray*}

\noindent The posterior pdf of $\alpha$ is not known, but the plot of its show that it is symmetric and close to normal distribution. So to generate random numbers from this distributions, we use the Metropolis-Hastings method with normal proposal distribution. Note that the Metropolis algorithm considers only symmetric proposal distributions. Therefore the algorithm of Gibbs sampling is as follows:
\begin{enumerate}
\item [1.] Start with an initial guess $( \alpha^{(0)}, \theta_1^{(0)}, \theta_2^{(0)} )$.
\item [2.] Set t = 1.
\item [3.] Using Metropolis-Hastings, generate $\alpha^{(t)}$ from $ f_\alpha=f(\alpha|\theta_1^{(t-1)},\theta_2^{(t-1)},data)$
with the $N(\alpha^{(t-1)},1)$ proposal distribution.
\item [4.] Generate $\theta_1^{(t)}$ from $ IG(r_1+a_1,b_1+\sum_{i=1}^{r_1}x_i^{\alpha^{(t-1)}} + (n-r_{1})u_{1}^{\alpha^{(t-1)}})$.
\item [5.] Generate $\theta_2^{(t)}$ from $ IG(r_2+a_2,b_2+\sum_{j=1}^{r_2}y_j^{\alpha^{(t-1)}} + (m-r_{2})u_{2}^{\alpha^{(t-1)}})$ .
\item [6.] Compute $R^{(t)}$ from \re{3}.
\item  [7.] Set $t = t + 1$.
\item [8.] Repeat steps 3-7, $M$ times.
\end{enumerate}

Note that in step {3}, we use the Metropolis-Hastings algorithm with $ q \sim N(\alpha^{(t-1)}, \sigma^{2}) $ proposal distribution as follows:\\
a. Let $ x = \alpha^{(t-1)}$.\\
b. Generate $  y $ from the proposal distribution q.\\
c. Let $ p(x, y) = \min\lbrace 1,\dfrac{f_{\alpha}(y)}{f_{\alpha}(x)}\dfrac{q(x)}{q(y)}\rbrace  $.\\
d. Accept $y$ with probability $ p(x, y) $ or accept $ x $ with probability $ 1- p(x, y)$.\\
Now the approximate posterior mean, and posterior variance of $R$
become
\begin{eqnarray*}
\widehat{E}(R|data)&=&\frac{1}{M}\sum_{t=1}^{M}R^{(t)},\\
\widehat{Var}(R|data)&=&\frac{1}{M}\sum_{t=1}^{M}\left(R^{(t)}-\widehat{E}(R|data)\right)^2.
\end{eqnarray*}

Based on $M$ and $R$ values, using the method proposed
by Chen and Shao (1999), a $100(1-\gamma)\%$ HPD credible interval can be constructed as $
\left(R_{[\frac{\gamma}{2}M]},R_{[(1-\frac{\gamma}{2})M]}\right)$, where $R_{[\frac{\gamma}{2}M]}$ and $R_{[(1-\frac{\gamma}{2})M]}$ are the $[\frac{\gamma}{2}M]$-th smallest integer and the $[(1-\frac{\gamma}{2})M]$-th smallest integer of $\{R_t,~t=1,2,\cdots,M\}$, respectively.

\vspace{2ex}

\section{Numerical Computations}
\paragraph{ }

In this section, analysis of a real data set and a Monte Carlo
simulation are presented to illustrate all the estimation methods
described in the preceding sections.

\subsection{Simulation Study }

In this subsection,  We  compare the performances of the MLE, AMLE
and the Bayes estimates with respect to the squared error loss
function in terms of biases, and mean squares errors (MSE). We also
compare different confidence intervals, namely the confidence
intervals obtained by using asymptotic distributions of the MLE, and
two different bootstrap confidence intervals and the HPD credible
intervals in terms of the average confidence lengths. For computing
the Bayes estimators and HPD credible intervals, we assume 2 priors
as follows:
\begin{eqnarray*}
&&\hspace{3cm}\mbox{Prior 1:}~~~a_j=0, ~~b_j=0,~~~j=1,2,3,\\
&&\hspace{3cm}\mbox{Prior 2:}~~~a_j=1, ~~b_j=2,~~~j=1,2,3.
\end{eqnarray*}
Prior 1 is the non-informative gamma prior for both the shape and
scale parameters. Prior 2 is informative gamma prior.

For different censoring schemes and different priors, we report the
average estimates, and MSE of the MLE, AMLE and Bayes estimates of R
over 1000 replications. The results are reported in Table 1. In our
simulation experiments for both the bootstrap methods, we have
computed the confidence intervals based on 250 re-sampling.
\newpage
\begin{center}
\scriptsize{\textbf{Table 1}.The average estimates(A.E) and MSE of the MLE, AMLE and Bayes estimates of $R$ when  $ m=n=30$ and $(\alpha,\theta_1,\theta_2)=(1.5, 1, 1)$ .\\
\vspace{.4 cm}
 \begin{tabular}{|c|c|c|c|c|cc|}

  \hline
 ( $r_{1}, T_{1}$)    &  ( $r_{2}, T_{2}$)  &   &  MLE  &  AMLE &  BS  &          \\
     \cline{6-7}
                          &    &  &     &      & prior 1 & prior 2     \\
  \hline
   & $(20, 1)$  & A.E  &0.4929   & 0.4921   & 0.5074  &   0.5067     \\
   &                 & MSE  & 0.0081  & 0.0093  &  0.0047  & 0.0031        \\
   \cline{2-7}
   & $(25, 1)$  & A.E  &  0.4941  & 0.5040  &  0.5068  & 0.5037      \\
 (20, 1)  &             & MSE  & 0.0072 & 0.0084  & 0.0035  & 0.0032           \\
  \cline{2-7}
   & $(20, 2)$  &  A.E &0.5049 &  0.4951 &0.5043  &  0.5039   \\
   &              & MSE    &0.0074&   0.0090&  0.0045 &   0.0039    \\
 \cline{2-7}
   & $(25, 2)$  &  A.E  & 0.4957 & 0.4953  &0.5049  &0.5036     \\
   &              & MSE    &0.0068 & 0.0071  & 0.0049  &    0.0041   \\
  \cline{2-7}
   & $(30, 2)$  &  A.E  &0.4986&  0.4979 &0. 5033 &  0. 5027      \\
   &              & MSE    &0.0065 &  0.0074 &0.0053 & 0.0037    \\
\hline

  & $(20, 1)$  &  A.E & 0.5058 & 0.4947  & 0.5051 &   0.5047     \\
   &              & MSE    &  0.0071 & 0.0078  &   0.0066 &     0.0061     \\
   \cline{2-7}
   & $(25, 1)$  &  A.E & 0.5045 &  0.5052 & 0.5038 & 0.5031    \\
   (25,1) &              & MSE    &  0.0066 & 0.0067  & 0.0054 &   0.0056      \\
  \cline{2-7}
   & $(20, 2)$  &  A.E  &0.5041& 0.5046  & 0.5039 &0.5022    \\
   &              & MSE    &0.0067 &  0.0069 &  0.0061&   0.0057    \\
 \cline{2-7}
   & $(25, 2)$  & A.E  &  0.5033  &  0.5039 & 0.5032  &  0.5023   \\
   &              & MSE    &  0.0058  &0.0061   & 0.0055    &    0.0046 \\
  \cline{2-7}
   & $(30, 2)$  &  A.E  & 0.4979& 0.4983  &   0.5024 &   0.5022  \\
   &              & MSE    &  0.0058 &   0.0049&   0.0049  & 0.0038  \\

\hline
 & $(20, 1)$  &  A.E  &  0.5035   &  0.5036   &  0.5033  &  0.5027      \\
   &              & MSE    & 0.0067  &  0.0074   &  0.0057  &    0.0043  \\
   \cline{2-7}
   & $(25, 1)$  &  A.E  & 0.4971 & 0.4956    &0.5029 &  0.5024   \\
(20,2)   &      & MSE    &  0.0054 &   0.0069   & 0.0046 &  0.0033       \\

  \cline{2-7}
    &$(20, 2)$ &  A.E    &0.5029   &  0.5031    &   0.5026  &  0.5025         \\
   &        & MSE   & 0.0066  &  0.0069   &  0.0048 &    0.0037         \\
 \cline{2-7}
   & $(25, 2)$  &  A.E  & 0.4979 & 0.4968 &0.5022  &0.5020       \\
   &              & MSE    &  0.0065& 0.0059  &  0.0043  &     0.0031      \\
  \cline{2-7}
   & $(30, 2)$  & A.E  &  0.4986 & 0.4977  & 0.5018  & 0.5016      \\
   &              & MSE    & 0.0056   & 0.0061   &    0.0039 &       0.0023     \\

\hline
   & $(20, 1)$  & A.E  &  0.4974 &  0.4959   & 0.5027   & 0.5019    \\
   &              & MSE    &  0.0062 &  0.0071 &  0.0045    &   0.0037      \\
   \cline{2-7}
   & $(25, 1)$  &  A.E  &0.4979&  0.4964  &  0.5020&0.5018  \\
 (25,2)  &              & MSE    & 0.0051& 0.0061   &  0.0043   &  0.0023      \\
  \cline{2-7}
   & $(20, 2)$ &  A.E  &  0.5023 & 0.4968   &  0.5021 & 0.5020          \\
   &              & MSE    &  0.0057 &  0.0070   &  0.0049   &  0.0027       \\
 \cline{2-7}
   & $(25, 2)$& A.E  & 0.5019   &0.4982  &0.5019   & 0.5017         \\
    &              & MSE    &0.0050    & 0.0054    &  0.0036  &   0.0021       \\
  \cline{2-7}
   & $(30, 2)$& A.E  & 0.4988  &0.4983   &0.5014   &0.5010          \\
   &              & MSE    & 0.0044 & 0.0041  &  0.0041  &    0.0013     \\

   \hline
     & $(20, 1)$  &  A.E  & 0.5047 & 0.4952   &  0.5042  & 0.5029    \\
   &              & MSE    & 0.0066  & 0.0076  &   0.0059  & 0.0043       \\
   \cline{2-7}
   & $(25, 1)$  &  A.E  & 0.4951 &  0.4959   & 0.5031   &0.5016  \\
 (30,2)  &              & MSE    &  0.0059 & 0.0064 & 0.0048   & 0.0032     \\
  \cline{2-7}
   & $(20, 2)$ &  A.E  &   0.5068 & 0.4953    &  0.5043 & 0.5021          \\
   &              & MSE    &  0.0064 &  0.0068    &0.0065   &    0.0029      \\
 \cline{2-7}
   & $(25, 2)$& A.E  &0.5034   &  0.5029  &0.5016   & 0.5009         \\
    &              & MSE    &0.0050    & 0.0057   &0.0047  &0.0014        \\
  \cline{2-7}
   & $(30, 2)$& A.E  &0.5005  &  0.5013  &0.5006   &0.5003          \\
   &              & MSE    & 0.0043  &  0.0049 &0.0051  & 0.0008       \\
 \hline
\end{tabular}}
\end{center}

\newpage
\begin{center}
\scriptsize{\textbf{Table 2}. Average confidence/credible length for estimators of $R$.\\
\vspace{.4cm}
 \begin{tabular}{|c|c|c|c|c|cc|}
  \hline
  ( $r_{1}, T_{1}$)  &  ( $r_{2}, T_{2}$)  &  MLE  &  Boot-t & Boot-p  &  BS  &          \\
     \cline{6-7}
      &    &     &        &        & prior 1 & prior 2    \\
  \hline
 & $(20, 1)$ &0.3854  & 0.4071 &   0.3966 &0.3717  & 0.3416 \\
\cline{2-7}
 &$(25, 1)$&  0.3779&  0.3988 &  0.3824 &  0.3714 & 0.3197        \\
 \cline{2-7}
(20, 1) &$(20, 2)$ & 0.3815  & 0.3874 & 0.3902 &  0.3759 &  0.2989     \\
 \cline{2-7}
  &$(25, 2)$&  0.3426&0.3747&0.3592&   0.3217&  0.2996    \\
 \cline{2-7}
 &$(30, 2)$ &  0.3140 &0.3706  &  0.3269&  0.2975 &0.2661    \\
\hline

 & $(20, 1)$ & 0.3785  & 0.3884 &  0.3877 & 0.3419  & 0.3055 \\
\cline{2-7}
 &$(25, 1)$&  0.3766 &  0.3899 &  0.3813 &  0.3124 & 0.2892     \\
 \cline{2-7}
 (25, 1) &$(20, 2)$&0.3711  & 0.3914 & 0.3732 &  0.3053 &  0.2971     \\
 \cline{2-7}
  &$(25, 2)$&0.3463 &0.3618&0.3557&   0.2902&  0.2566   \\
 \cline{2-7}
 &$(30, 2)$& 0.2939 &0.3314  &  0.3273&  0.2613 &0.2387   \\
\hline

  & $(20, 1)$ & 0.3461  & 0.3597 &  0.3444 & 0.3143  & 0.2819      \\
\cline{2-7}
 &$(25, 1)$&0.3211  & 0.3424 &0.3229  &  0.3078 &   0.2413    \\
 \cline{2-7}
(20, 2) &$(20, 2)$&0.3357  & 0.3527 & 0.3289 &  0.3053 &  0.2673    \\
 \cline{2-7}
  &$(25, 2)$&0.3103&0.3315&0.3071&   0.2642&  0.2374     \\
 \cline{2-7}
 &$(30, 2)$ & 0.2846 &0.3363  &  0.2978&  0.2560 &0.1956       \\
\hline

  & $(20, 1)$ &  0.3484&   0.3677& 0.3565 &  0.2835   &  0.2634     \\
\cline{2-7}
 &$(25, 1)$& 0.3192  & 0.3275 &0.2905  &  0.2764 &  0.2312     \\
 \cline{2-7}
(25, 2) &$(20, 2)$   &  0.3248    &  0.3350  &  0.3132   &   0.2714  &  0.2276   \\
 \cline{2-7}
  &$(25, 2)$  &  0.2925  &   0.2850  & 0.3013    &    0.2558   &  0.2170    \\
 \cline{2-7}
 &$(30, 2)$ &  0.2614  &    0.2831  &  0.2738     &   0.2203   &   0.2111  \\
\hline

  & $(20, 1)$ &0.3417 & 0.3605&0.3452   & 0.2304   &   0.2079    \\
\cline{2-7}
 &$(25, 1)$& 0.3311& 0.3225   & 0.3308 & 0.2244 &  0.1716     \\
 \cline{2-7}
(30, 2) &$(20, 2)$ & 0.3255  & 0.3600&  0.3396 & 0.2372  &   0.1948   \\
 \cline{2-7}
  &$(25, 2)$ &  0.2779   & 0.2813 &  0.2800  & 0.2052  &   0.1620    \\
 \cline{2-7}
 &$(30, 2)$  &  0.2462    &   0.2804   &   0.2748 & 0.2034   &   0.0913    \\
\hline
\end{tabular}}
\end{center}
\bigskip

\subsection{Data Analysis}

Here we present a data analysis of the strength data
reorted by Badar and Priest (1982). This data, represent the
strength measured in GPA for single carbon fibers, and impregnated
1000-carbon fiber tows. Single fibers were tested under tension at
gauge lengths of 20mm (Data Set 1) and 10mm (Data Set 2). These data
have been used previously by Raqab and Kundu (2005), Kundu and Gupta
(2006), Kundu and Raqab (2009) and Asgharzadeh et al (2011). The data are presented in Tables
3 and 4.
\begin{center}
\scriptsize{
\textbf{Table 3}. Data Set 1 (gauge lengths of 20 mm).\\
\vspace{.2cm}
\begin{tabular}{cccccccccc}
\hline
1.312 & 1.314 & 1.479 & 1.552 & 1.700 & 1.803 & 1.861 & 1.865 & 1.944 & 1.958 \\
1.966 & 1.997 & 2.006 & 2.021 & 2.027 & 2.055 & 2.063 & 2.098 & 2.140 & 2.179 \\
2.224 & 2.240 & 2.253 & 2.270 & 2.272 & 2.274 & 2.301 & 2.301 & 2.359 & 2.382 \\
2.382 & 2.426 & 2.434 & 2.435 & 2.478 & 2.490 & 2.511 & 2.514 & 2.535 & 2.554 \\
2.566 & 2.570 & 2.586 & 2.629 & 2.633 & 2.642 & 2.648 & 2.684 & 2.697 & 2.726 \\
2.770 & 2.773 & 2.800 & 2.809 & 2.818 & 2.821 & 2.848 & 2.880 & 2.954 & 3.012 \\
3.067 & 3.084 & 3.090 & 3.096 & 3.128 & 3.233 & 3.433 & 3.585 & 3.585\\
\hline
\end{tabular}}
\end{center}

\newpage

\begin{center}
\scriptsize{
\textbf{Table 4}. Data Set 2 (gauge lengths of 10 mm).\\
\vspace{.2cm}
\begin{tabular}{cccccccccc}
\hline
1.901 & 2.132 & 2.203 & 2.228 & 2.257 & 2.350 & 2.361 & 2.396 & 2.397 & 2.445 \\
2.454 & 2.474 & 2.518 & 2.522 & 2.525 & 2.532 & 2.575 & 2.614 & 2.616 & 2.618 \\
2.624 & 2.659 & 2.675 & 2.738 & 2.740 & 2.856 & 2.917 & 2.928 & 2.937 & 2.937 \\
2.977 & 2.996 & 3.030 & 3.125 & 3.139 & 3.145 & 3.220 & 3.223 & 3.235 & 3.243 \\
3.264 & 3.272 & 3.294 & 3.332 & 3.346 & 3.377 & 3.408 & 3.435 & 3.493 & 3.501 \\
3.537 & 3.554 & 3.562 & 3.628 & 3.852 & 3.871 & 3.886 & 3.971 & 4.024 & 4.027 \\
4.225 & 4.395 & 5.020\\
\hline
\end{tabular}}
\end{center}

Kundu and Gupta (2006) analyzed these data sets using two-parameter
Weibull distribution after subtracting 0.75 from both these data
sets. After subtracting 0.75 from all the points of these data sets,
Kundu and Gupta (2006)  observed that the Weibull distributions with
equal shape parameters fit to both these data sets. Based on the complete data, we plot the histogram of the samples of $ \alpha $ generated by MCMC method along with their exact
posterior density function in Figure 1. From the Figure 1 it is clear that the exact posterior density function match quite well with the simulated samples obtained using MCMC method.\

 For illustrative purposes, we havegenerated two different hybrid censored samples from above data sets after subtracting 0.75:\

\noindent \textbf{Scheme 1}: $r_{1} = 45, T_{1} =2.5, r_{2} = 40, T_{2} =2.5$.\

The MLE and AMLE of $R $ become 0.3958 and 0.3872; and the corresponding 95\%
confidence intervals become (0.2723, 0.5193) and (0.2617, 0.4891), respectively. We also
obtain the 95\% Boot-p and Boot-t confidence intervals as (0.3045, 0.5523) and (0.3106, 0.5526), respectively.\

For the Bayesian estimate of $ R $, we use non-proper priors on
$ \theta_{1} $, $ \theta_{2} $ and $ \alpha $, i.e., $ a_{1} = a_{2}= a_{3}= b_{1} = b_{2} =b_{3} = 0$.
Based on the above priors, we obtain 0.3933 as the Bayes estimate
of $ R $, under the squared error loss function. We also compute
the 95\%  highest posterior density (HPD) credible interval
of $ R $ as (0.2937, 0.4829).\

\begin{figure}
\centering
\includegraphics [height=9cm,width=.7\textwidth]{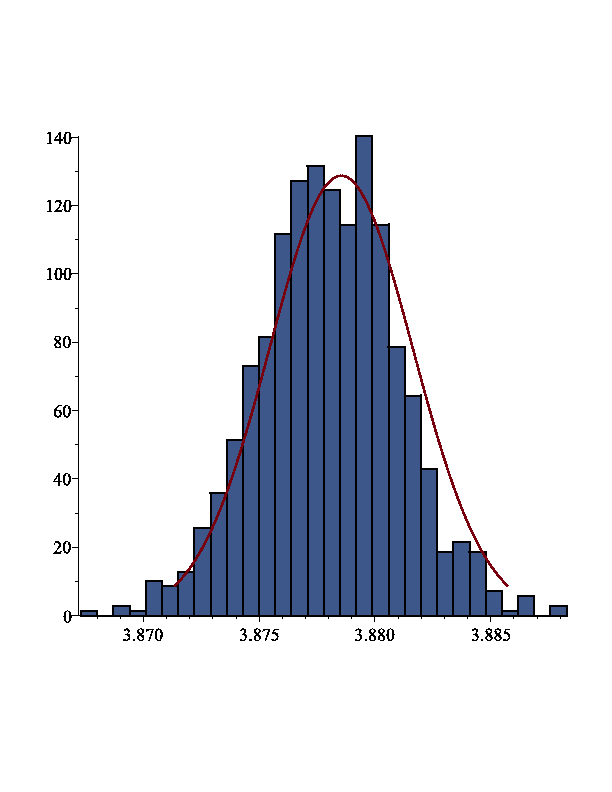}
\caption{ Posterior density function of $  \alpha$.}
\end{figure}

\noindent \textbf{Scheme 2}: $r_{1} = 35, T_{1} =1.7, r_{2} = 25, T_{2} =2.2$.\

In this case, the MLE, AMLE and Bayes estimations of $R$
become 0.4024, 0.4092 and 0.4238; and the corresponding 95\% confidence interval’s become (0.2781, 0.5267), (0.2960, 0.5531) and (0.2870, 0.5371) respectively. We also obtain the 95\% Boot-p and Boot-t confidence intervals as (0.3359, 0.5711) and (0.3484, 0.5964) respectively.

\section{Conclusions}

In this paper we have addressed the inferential issues on $R = P (X > Y)$, when $X$ and $Y$ are independent and the have Weibull
distribution with the same shape but different scale parameters.  It is assumed that the data are hybrid censored on both $X$ and
$Y$.  We have provided the MLE of $R$, and it is observed that the MLE of $R$ cannot be obtained in explicit form.  But it can be
obtained by solving a one dimensional non-linear equation.  We have also provided the AMLE of $R$, and it can be obtained explicitly.
Extensive Monte Carlo simulations indicate that the performances of  MLE and AMLE are very similar, hence for all practical purposes
AMLE can be used in practice.  We have further considered the Bayesian inference on $R$ based on fairly general inverted gamma priors
on the scale parameters, and an independent gamma prior on the common shape parameter.  The Bayes estimator cannot be obtained in
explicit form, we have Gibbs sampling technique to compute the Bayes estimate and also to compute the credible interval.  It is observed
that the performances of the Bayes estimators are very satisfactory, and if some prior informations are available on the unknown
parameters,  Bayes estimator should be prefered compared to MLE or AMLE, as expected.

\vspace{.2cm}
\begin{center}
\section*{Appendix }
\end{center}
\textbf{AMLE of $R$ for other cases}:

When $u_1=T_{1}$ and $u_2=T_{2}$, we expand  $p(z_i)$, and $q(u_1^*)$ in Taylor series
around the points $\mu_i$ and $\mu_{r_1}^*$, respectively. Further, we also expand the termes $\bar{p}{(w_j)}$, and $\bar{q}(u_2^*)$ around the points $\mu_j$ and $\mu_{r_2}^*$, respectively .
Similarly, considering only the first order derivatives and neglecting the higher order derivatives, the AMLEs of $\alpha$, $\theta_1$ and $\theta_2$,  can be obtained as
\[
\tilde{\mu}_1=A_1+B_1\tilde{\sigma},\;\;\tilde{\mu}_2=A_2+B_2\tilde{\sigma},\;\;\mbox{and}\;\;
\tilde{\sigma}=\frac{-D+\sqrt{D^2-4(R_1+R_2)E}}{2(R_1+R_2)},
\]
\noindent where

\begin{eqnarray*}
A_1=\frac{\sum_{i=1}^{r_1}\beta_i t_i+(n-r_1)\beta_{r_1^*} \ln(T_1)}{\sum_{i=1}^{r_1}\beta_i+(n-r_1)\beta_{R_1^*}},
&& B_1=\frac{\sum_{i=1}^{r_1}\alpha_i -(n-r_1)(1-\alpha_{r_1^*})}{\sum_{i=1}^{r_1}\beta_i+(n-r_1)\beta_{r_1^*}},
\end{eqnarray*}

\begin{eqnarray*}
A_2=\frac{\sum_{j=1}^{r_2}\beta_j s_j+(m-r_2)\beta_{r_2^*}s_{r_2}}{\sum_{j=1}^{r_2}\beta_j+(m-r_2 )\beta_{r_2^*}},&& B_2=\frac{\sum_{j=1}^{r_2}\alpha_j -(m-r_2)(1-\alpha_{r_2^*})}{{\sum_{j=1}^{r_2}\beta_j+(m-r_2)\beta_{r_2^*}}},
\end{eqnarray*}

\begin{eqnarray*}
D&=&\sum_{i=1}^{r_1}\alpha_i(\ t_i -3A_1)-(n-r_1)(1-\alpha_{r_1^*}) (\ln(T_1)-3A_1)+2A_1B_1
\sum_{i=1}^{r_1}\beta_i +2A_1 B_1(n-r_1)\beta_{r_1^*}\\
 &+&\sum_{j=1}^{r_2}\alpha_j(\ s_j -3A_2)-(m-r_2)(1-\alpha_{r_2^*})(\ln(T_2)-3A_2)+2A_2B_2\sum_{j=1}^{r_2}\beta_i+2A_2 B_2(m-r_2)\beta_{r_2^*},\\
\\
E&=&\sum_{i=1}^{r_1}\beta_i t_i(t_i-A_1)+(n-r_1)\beta_{r_1^*}(\ln(T_1))^2-(n-r_1)A_1\beta_{r_1^*}\ln(T_1)
+\sum_{j=1}^{r_2}\beta_j s_j(s_j-A_2) \hspace{1cm}\\
&+&(m-r_2)\beta_{r_2^*}(\ln(T_2))^2 -(m-r_2)A_2\beta_{R_2^*}\ln(T_2),
\end{eqnarray*}

\noindent here
\begin{eqnarray*}
 \alpha_{r_i^*} &=& 1+\ln(q_{r_i^*})(1-\ln(-\ln (q_{r_i^*}))), \hspace{1cm}\beta_{r_i^*}=\ln (q_{r_i^*}) , i=1,2.
\end{eqnarray*}

 Therefore, the approximate MLE of $R$ is given by
$$
\tilde{R}=\frac{\tilde{\theta}_1}{\tilde{\theta}_1+\tilde{\theta}_2},
$$
\noindent where
$$
\tilde{\theta}_1=\exp{\left(\frac{1}{\tilde{\sigma}}(A_1+B_1\tilde{\sigma})\right)},\;\;
\mbox{and}\;\;\tilde{\theta}_2=\exp{\left(\frac{1}{\tilde{\sigma}}(A_2+B_2\tilde{\sigma})\right)}.
$$
Similarly, AMLE of $R$ for other cases(when $u_1=T_{1}$, $u_2=x_{R_2}$ or $u_1=x_{R_1}$, $u_2=T_{2}$ )can be obtained.

\end{document}